\documentclass[pra,superscriptaddress,twocolumn]{revtex4}

\usepackage[dvips]{graphics}
\usepackage{amssymb}
\usepackage{amsmath}
\usepackage{epsfig}
\usepackage{latexsym}
\usepackage{color}
\usepackage{rotating}

\long\def\symbolfootnote[#1]#2{\begingroup
\def\thefootnote{\fnsymbol{footnote}}\footnote[#1]{#2}\endgroup}

\begin{document}

\title{Signatures of Hong-Ou-Mandel Interference at Microwave Frequencies}

\author{M. J. Woolley\footnote{Current address: School of Engineering and Information Technology, UNSW Canberra (Australian Defence Force Academy), Australian Capital Territory, 2600, Australia.}} 
\affiliation{Depart\'{e}ment de Physique, Universit\'{e} de Sherbrooke, Sherbrooke, QC J1K 2R1, Canada}
\author{C. Lang}
\affiliation{Department of Physics, ETH Z\"{u}rich, Z\"{u}rich, CH-8093, Switzerland}
\author{C. Eichler}
\affiliation{Department of Physics, ETH Z\"{u}rich, Z\"{u}rich, CH-8093, Switzerland}
\author{A. Wallraff}
\affiliation{Department of Physics, ETH Z\"{u}rich, Z\"{u}rich, CH-8093, Switzerland}
\author{A.Blais}
\affiliation{Depart\'{e}ment de Physique, Universit\'{e} de Sherbrooke, Sherbrooke, QC J1K 2R1, Canada}

\begin{abstract}
\noindent
Two-photon quantum interference at a beam splitter, commonly known as Hong-Ou-Mandel interference, was recently demonstrated with \emph{microwave-frequency} photons by Lang \emph{et al.}\,\cite{lang:microwaveHOM}. This experiment employed circuit QED systems as sources of microwave photons, and was based on the measurement of second-order cross-correlation and auto-correlation functions of the microwave fields at the outputs of the beam splitter. Here we present the calculation of these correlation functions for the cases of inputs corresponding to: (i) trains of \emph{pulsed} Gaussian or Lorentzian single microwave photons, and (ii) resonant fluorescent microwave fields from \emph{continuously-driven} circuit QED systems. The calculations include the effects of the finite bandwidth of the detection scheme. In both cases, the signature of two-photon quantum interference is a suppression of the second-order cross-correlation function for small delays. The experiment described in Ref.~\onlinecite{lang:microwaveHOM} was performed with trains of \emph{Lorentzian} single photons, and very good agreement between the calculations and the experimental data was obtained. 
\end{abstract}

\maketitle

\section{Introduction}
Hong-Ou-Mandel interference is a two-photon quantum interference effect whereby two \emph{indistinguishable} photons incident at either port of a balanced beam splitter will always be detected to coalesce at one or the other output port of the beam splitter, but never with one photon at each output port. This interference effect diminishes as the distinguishability of the photons is increased. It was first demonstrated by Hong, Ou and Mandel in 1987 using photons produced via parametric down-conversion \cite{HOM}. The importance of this effect in optical implementations of quantum information processing schemes was realized later \cite{kimble,KLM}, motivating further demonstrations. They have been achieved both by using a single source \cite{santori,legero2} and by using two independent sources \cite{deRiedmatten,beugnon,maunz} of photons. In addition to these \emph{pulsed} interference experiments, two-photon quantum interference using \emph{continuously-driven} sources of resonant fluorescent light has also been demonstrated \cite{blatt}.

However, until the recent experiment of Lang \emph{et al.}\,\cite{lang:microwaveHOM}, a comparable demonstration of the same effect with \emph{microwave-frequency} photons had been lacking. This experiment was performed using a circuit QED system \cite{wallraff,blais}, configured as pulsed sources of single microwave photons \cite{bozyigit}. Figs.~2(c)-(f) of Ref.~\onlinecite{lang:microwaveHOM} show both the measured and calculated second-order correlation functions of the microwave fields at the output of the beam splitter. The calculation of these theory curves, corresponding to inputs consisting of trains of Lorentzian photons, is presented here, along with the same calculation for Gaussian photons. One could also perform the experiment with continuously-driven circuit QED systems, operated in the regime of \emph{resonant photon blockade} \cite{resonantphotonblockade}. We also calculate the output second-order correlation functions in this case, though these are not directly relevant to the experiments described in Ref.~\onlinecite{lang:microwaveHOM}. 
%both without and with a polarisation degree of freedom. 

The calculation of the correlation functions in the pulsed case proceeds in a manner similar to that of Legero \emph{et al.}\,\cite{legero1}, undertaken to describe their experiments on \emph{time-resolved} two-photon quantum interference \cite{legero2}. We generalise their calculations in a number of ways. First, we consider Lorentzian (in addition to Gaussian) single photons. This is because here the spatio-temporal properties of the emitted photons are determined purely by the decay rate of the circuit QED source, rather than by additional broadening mechanisms. Second, we calculate the correlation functions for a \emph{train} of single photons, leading to features in the calculated correlation functions at (and around) integer multiples of the pulse period. Further, these correlation functions are evaluated as \emph{circular} correlation functions, meaning that the correlations for a finite pulse train are obtained by wrapping the time index around the finite sequence of pulses. Third, in order to make a comparison with experimental data, we subject the calculated correlation functions to filtering representative of the filtering that the measured data is subjected to. 

The second-order correlation functions at the output of the beam splitter are evaluated, in terms of correlation functions of input fields, in Sec.~\ref{sec:correlations}. These are explicitly evaluated, for both individual and trains of pulsed single-photons (Gaussian and Lorentzian), in Sec.~\ref{sec:singlephotons}. The correlation functions for continuously-driven sources, both without and with polarisation degrees of freedom, are evaluated in Sec.~\ref{sec:continuouslydrivensources}. Note that neither the details of the extraction of the microwave correlation functions from the directly measured correlation functions in Ref.~\onlinecite{lang:microwaveHOM}, nor the details of the quantum state tomography employed in Ref.~\onlinecite{lang:microwaveHOM}, are described here. They have been discussed elsewhere \cite{silva,eichler}. 

\section{Output Second-Order Correlation Functions }
\label{sec:correlations}

\subsection{Intensity Correlation Functions}
Suppose that the two fields incident on the input ports of the beam splitter are described by dimensionless, time-dependent quantum fields denoted by $\hat{a}'(t)$ and $\hat{b}'(t)$. %The experimental layout is depicted in Fig.~\ref{fig:exptlayout}. 
Unitarity of the (balanced) beam splitter implies that the output fields, $\hat{a}(t)$ and $\hat{b}(t)$, are then given by the Heisenberg picture evolution \cite{quantumoptics},
\begin{equation}
\left[ \begin{array}{c} \hat{a}(t) \\ \hat{b}(t) \end{array} \right] = \frac{1}{\sqrt{2}} \left[ \begin{array}{cc} 1 & -1 \\ 1 & 1 \end{array} \right] \left[ \begin{array}{c} \hat{a}'(t) \\ \hat{b}'(t) \end{array} \right] . \label{eq:BS}
\end{equation}
We are interested, primarily, in the statistics of the photons output from the beam splitter. These are described by correlation functions of the number operators of the output modes, the number operators being denoted by $\hat{n}_c (t) \equiv \hat{c}^\dagger (t) \hat{c} (t)$. Then, the quantities we seek are the \emph{normally-ordered} correlation functions of the output mode number operators:
\begin{equation}
\left\langle : \hat{n}_a (t+\tau ) \hat{n}_{r} (t) : \right\rangle = \left\langle \hat{r}^\dagger (t) \hat{a}^\dagger (t+\tau ) \hat{a} (t+\tau ) \hat{r} (t) \right\rangle , \label{eq:CFs}
\end{equation}
with $\hat{r}$ representing either $\hat{a}$ (for the output intensity \emph{auto-correlation} function) or $\hat{b}$ (for the output intensity \emph{cross-correlation} function). These correlation functions allow one to discriminate between quantum and classical fields~\cite{quantumoptics}. Since the intensity of a field is proportional to the number of photons it contains, henceforth we shall refer to correlation functions of the form of Eq.~(\ref{eq:CFs}) as \emph{intensity} correlation functions. The term \emph{second-order correlation function} shall be reserved here for functions of the difference in detection times ($\tau$) alone, to be introduced below. The correlation functions $\langle : \hat{n}_b(t+\tau ) \hat{n}_r (t) : \rangle$ give the same results due to the symmetry of Eq.~(\ref{eq:BS}). 

Clearly, we can obtain the intensity correlation functions of Eq.~(\ref{eq:CFs}) in terms of the input fields by substituting the expressions in Eq.~(\ref{eq:BS}). The result includes sixteen fourth-order moments of the time-dependent input fields. 
Now, if it is assumed that the sources produce separable (i.e.\,not entangled) fields (clearly a reasonable assumption for independent sources), the four-term cross-correlation functions of input fields may each be decomposed into products of two auto-correlation functions of the input fields. Therefore, the output intensity auto-correlation and cross-correlation functions are:
\begin{widetext}
\begin{eqnarray}
\left\langle : \hat{n}_a (t+\tau ) \hat{n}_{a(b)} (t) : \right\rangle & = & \frac{1}{4} \sum_{ (\hat{c}, \hat{d} ) \in \Pi (\hat{a}', \hat{b}' )  } \left[ \langle \hat{c}^\dagger (t) \hat{c}^\dagger (t+\tau ) \hat{c} (t+\tau ) \hat{c} (t) \rangle + \left\langle \hat{c}^\dagger (t) \hat{c} (t) \right\rangle \langle \hat{d}^\dagger (t+\tau ) \hat{d} (t+\tau ) \rangle \right. \nonumber \\
& & \ \ \ \ \ \ \ \ \  \ \ \ \ \  \pm \left\langle \hat{c}^\dagger (t+\tau) \hat{c} (t) \right\rangle \langle \hat{d}^\dagger (t ) \hat{d} (t+\tau ) \rangle \pm \left\langle \hat{c}^\dagger (t+\tau) \hat{c}^\dagger (t) \right\rangle \langle \hat{d} (t+\tau ) \hat{d} (t ) \rangle \nonumber \\
& & \ \ \ \ \ \ \ \ \ \ \ \ \ \ + \langle \hat{d}(t) \rangle \left\langle \hat{c}^\dagger (t) \hat{c}^\dagger (t+\tau ) \hat{c}(t+\tau ) \right\rangle + \langle \hat{d}^\dagger (t) \rangle \left\langle  \hat{c}^\dagger (t+\tau ) \hat{c}(t+\tau ) \hat{c} (t) \right\rangle \nonumber \\
& & \ \ \ \ \ \ \ \ \ \ \ \ \ \ \left. \pm \langle \hat{d}(t+\tau) \rangle \left\langle  \hat{c}^\dagger (t+\tau ) \hat{c}^\dagger (t ) \hat{c} (t) \right\rangle \pm \langle \hat{d}^\dagger (t+\tau) \rangle \left\langle  \hat{c}^\dagger (t ) \hat{c} (t+\tau ) \hat{c} (t) \right\rangle \right] , \label{eq:evaluatedNumberCF}
\end{eqnarray}
\end{widetext}
where the $+$ and $-$ signs correspond to the auto-correlation and cross-correlation functions, respectively. Note that the summation is over the set of permutations of the input mode operators $\hat{a}'$ and $\hat{b}'$; that is, $\Pi ( \hat{a}', \hat{b}' ) \equiv \{ (\hat{a}', \hat{b}'), ( \hat{b}', \hat{a}') \} $. The output intensity correlation functions depend on the input intensity correlation functions [first term in Eq.~(\ref{eq:evaluatedNumberCF})], the products of the intensities of the two inputs (second term), the products of the first-order correlation functions of the two inputs (third term), and on a number of \emph{phase-dependent} correlation functions of the inputs (terms four to eight). 

\subsection{Phase-Dependent Moments}
\label{sec:phasedependentmoments}
For the purpose of demonstrating Hong-Ou-Mandel interference, the contributions of the fourth through to the eighth terms to the intensity correlation functions of Eq.~(\ref{eq:evaluatedNumberCF}) are neither necessary nor desirable. We will consider their contribution using the example of the fourth term, though the same arguments may be applied to the fifth through to the eighth terms. Suppose that the fields corresponding to the input modes, $c$ and $d$ in Eq.~(\ref{eq:evaluatedNumberCF}), can be written as slowly-varying fields modulating oscillations at \emph{carrier frequencies} centred on $\omega_0$ and with a difference $\Delta$, and with some relative phase offset $\theta $: $\hat{c}(t) \equiv \bar{c}(t) e^{+i (\omega_{0}+\Delta/2 ) t + i\theta }$ and $\hat{d}(t) \equiv \bar{d}(t) e^{+i (\omega_{0}-\Delta /2 ) t} $. 
Accordingly, we can rewrite the fourth term in Eq.~(\ref{eq:evaluatedNumberCF}) as
\begin{eqnarray}
& & \left\langle \hat{c}^\dagger (t+\tau ) \hat{c}^\dagger (t) \right\rangle \langle \hat{d}(t+\tau ) \hat{d} (t) \rangle \nonumber \\
& = & \left\langle \bar{c}^\dagger (t+\tau ) \bar{c}^\dagger (t) \right\rangle \left\langle \bar{d}(t+\tau ) \bar{d} (t) \right\rangle e^{- i \Delta (2t+\tau ) } e^{-2i\theta} . \label{eq:phasedependentterm}
\end{eqnarray}
If $\Delta$ is non-zero this expression will remain oscillatory even in the limit $t \rightarrow \infty$. However, if $\left| \Delta \right|$ is large compared with the inverse time-scales of the features of interest in the input fields, this expression will average to zero over the time-scales of interest. Alternatively, for single-photon input states the phase-dependent second moments will always evaluate to zero. Even if the expression in Eq.~(\ref{eq:phasedependentterm}) does not automatically vanish, its effect can be averaged to zero over multiple runs of the experiment if the relative phase $\theta$ is uniformly distributed. In all cases of interest to us, the phase-dependent moments will vanish and the expression in Eq.~(\ref{eq:evaluatedNumberCF}) reduces to: 
\begin{eqnarray}
& & \overline{ \left\langle : \hat{n}_a (t+\tau ) \hat{n}_{a(b)} (t) : \right\rangle } \nonumber \\
& = & \frac{1}{4} \sum_{ ( \hat{c}, \hat{d} ) \in \Pi ( \hat{a}', \hat{b}' ) } \left[ \left\langle \hat{c}^\dagger (t) \hat{c}^\dagger (t+\tau ) \hat{c} (t+\tau ) \hat{c} (t) \right\rangle  \right. \nonumber \\
& & \ \ \ \ \ \ \ \ \ + \left\langle \hat{c}^\dagger (t) \hat{c} (t) \right\rangle \langle \hat{d}^\dagger (t+\tau ) \hat{d} (t+\tau ) \rangle \nonumber \\
& & \ \ \ \ \ \ \ \ \ \pm \left\langle \hat{c}^\dagger (t+\tau) \hat{c} (t) \right\rangle \langle \hat{d}^\dagger (t ) \hat{d} (t+\tau ) \rangle ] . \label{eq:CFEvenMoments}
\end{eqnarray}
Here the bar over the correlation function indicates that it is to be evaluated over many runs of the experiment. Henceforth, this shall be assumed for the correlation functions quoted and we drop this notation. The output intensity correlation functions are \emph{completely determined} by the input intensity correlation functions and products of the input first-order correlation functions. For the output intensity \emph{cross-correlation} function, the two-photon interference is seen in the cancellation of the second and third terms at $\tau = 0$ in Eq.~(\ref{eq:CFEvenMoments}). On the other hand, the second and third terms will add constructively for the output \emph{auto-correlation} function at $\tau = 0$, corresponding to photon coalescence. Whether or not this constitutes Hong-Ou-Mandel interference in the usual sense is determined by the input intensity correlation functions in the first term of Eq.~(\ref{eq:CFEvenMoments}). Although the preceding statements are true in general, the \emph{signature} of the interference for non-zero $\tau$ will depend on the form of the input fields.  

\subsection{Second-Order Correlation Functions}
The output intensity correlation functions, as expressed in Eqs.~(\ref{eq:evaluatedNumberCF}) and (\ref{eq:CFEvenMoments}), are explicitly dependent on two detection times, $t$ and $t+\tau$. The \emph{second-order correlation functions}, giving the likelihood of detecting a photon at some time $\tau$ assuming detection at time zero, may be obtained either by integrating or by taking a limit:
\begin{eqnarray}
G^{(2)}_{cd} (\tau ) & = & \int dt \left\langle : \hat{n}_c (t+\tau ) \hat{n}_d (t ) : \right\rangle \ \ \ {\rm or } \nonumber \\
G^{(2)}_{cd} (\tau ) & = & \lim_{t \rightarrow \infty} \left\langle : \hat{n}_c (t+\tau ) \hat{n}_d (t ) : \right\rangle ,  \label{eq:numberCF} 
\end{eqnarray}
respectively. Whether one integrates or takes a limit depends on whether the system's evolution is described in a (non-rotating) laboratory or a rotating frame. The quantities in Eq.~(\ref{eq:numberCF}) are conventionally referred to as \emph{second-order correlation functions} \cite{quantumoptics}. Similarly, we introduce the \emph{first-order correlation functions}:
\begin{eqnarray}
G^{(1)}_c (\tau ) & = & \int  dt \left\langle \hat{c}^\dagger (t) \hat{c} (t+\tau ) \right\rangle \ \ \ {\rm or} \nonumber \\
G^{(1)}_c (\tau ) & = & \lim_{t \rightarrow \infty}  \left\langle \hat{c}^\dagger (t) \hat{c} (t+\tau ) \right\rangle .
\end{eqnarray}
In the case where the second-order correlation functions are obtained by taking limits, the output second-order correlation functions follow immediately from Eq.~(\ref{eq:CFEvenMoments}) as:
\begin{eqnarray} 
G^{(2)}_{aa(ab)}( \tau ) & = & \frac{1}{4} \sum_{(c,d) \in \Pi (a',b')} \left\{ G^{(2)}_{c} (\tau ) + G^{(1)}_{c}(0) G^{(1)}_{d} (0) \right. \nonumber \\
& & \left.  \pm \left[ G^{(1)}_{c} (\tau ) \right]^* G^{(1)}_{d} (\tau )  \right\}, \label{eq:crossCorrTrain} 
\end{eqnarray}
where we have introduced the short-hand notation $G^{(2)}_c (\tau ) \equiv G^{(2)}_{cc} (\tau )$. 
If both sources exhibit identical statistics, meaning $G^{(1)}_{a'} (\tau ) = G^{(1)}_{b'} (\tau ) \equiv G^{(1)}_i (\tau ) $ and $G^{(2)}_{a'} (\tau ) = G^{(2)}_{b'} (\tau )  \equiv G^{(2)}_i (\tau )$, the output second-order correlation functions are simply:
\begin{equation} 
G^{(2)}_{aa(ab)} (\tau ) = \frac{1}{2} \left\{ G^{(2)}_i (\tau ) \pm \left| G^{(1)}_i (\tau ) \right|^2 + \left[ G^{(1)}_i(0) \right]^2 \right\} . \label{eq:CFsIO}
\end{equation}
As for Eq.~(\ref{eq:CFEvenMoments}), the second and third terms of the second-order cross-correlation functions cancel at $\tau = 0$, corresponding to photon interference at the beam splitter. We now consider the evaluation of these correlation functions for particular forms of the input fields. 

\section{Pulsed Single-Photon Sources}
\label{sec:singlephotons}

\subsection{Spatio-temporal Modes and Correlation Functions}
The case relevant to the experiment described in Ref.~\onlinecite{lang:microwaveHOM} is that in which each source produces a highly-pure, microwave-frequency single-photon state. We shall evaluate the correlation functions of Eq.~(\ref{eq:CFEvenMoments}) by introducing \emph{spatio-temporal mode functions} to describe these single photons. First, the time-dependent quantum fields introduced above, $\hat{c}(t)$ in general, may be represented in the frequency domain as $\hat{c}(\omega ) = \frac{1}{\sqrt{2\pi}} \int dt \ e^{+i\omega t} \hat{c} (t) $.
Then a single-photon pulse is conventionally represented using the \emph{spectral density} $\Phi (\omega )$; that is,
\begin{equation}
\left| 1 \right\rangle = \int d\omega \ \Phi (\omega ) \hat{c}^\dagger (\omega ) \left| 0 \right\rangle . \label{eq:SinglePhoton}
\end{equation}
One can show, by considering the action of $\hat{c}(t)$ on the state $\left| 1 \right\rangle$, that the time-dependent quantum field $\hat{c}(t)$ may be represented as
\begin{equation}
\hat{c}(t) = \zeta (t) \hat{c}_0, \ \ \ {\rm where} \ \ \ \zeta (t) = \frac{1}{\sqrt{2\pi}} \int d\omega \ \Phi (\omega ) e^{-i\omega t} 
\end{equation}
is the spatio-temporal mode function of the single-photon pulse, and $\hat{c}_0$ is the annihilation operator corresponding to this spatio-temporal mode. It is then clear that $\left\langle \hat{c}^\dagger (t) \hat{c} (t) \right\rangle = \left| \zeta (t) \right|^2 $, giving the probability of the photon being detected at time $t$. 

In the experiment of Ref.~\onlinecite{lang:microwaveHOM}, there are two independent microwave-frequency sources producing highly-pure single-photon states. Accordingly, we can write the beam splitter input fields (from each source) in terms of spatio-temporal mode functions as
\begin{equation}
\hat{a}' (t) = \zeta_a (t) \hat{a}'_0 , \ \ \ \hat{b}' (t) = \zeta_b (t) \hat{b}'_0 .
\end{equation}
Accordingly, the two-mode state input to the beam splitter may be written as $\hat{a}'^\dagger (t) \hat{b}'^\dagger (t) \left| 0 \right\rangle$. From Eq.~(\ref{eq:CFEvenMoments}), the output intensity correlation functions are:
\begin{eqnarray}
& & \left\langle : \hat{n}_a (t+\tau ) \hat{n}_{a(b)} (t) : \right\rangle \nonumber \\
& & \ \ \ = \frac{1}{4} \left| \zeta_a (t+\tau ) \zeta_b (t) \pm \zeta_b (t+\tau ) \zeta_a (t) \right|^2 . \label{eq:ExplicitCFs}
\end{eqnarray}
Here, as in Eq.~(\ref{eq:CFEvenMoments}), the potential for cancellation of the terms on the right-hand-side in the case of the intensity \emph{cross-correlation} function is clear. Indeed, at $\tau = 0$, we see that the intensity cross-correlation function vanishes irrespective of the precise form of the spatio-temporal mode functions. Since there are now no additional contributions to the output intensity cross-correlation functions, the intensity correlation function vanishes at $\tau = 0$. This suppression is the signature of Hong-Ou-Mandel interference. On the other hand, the terms in Eq.~(\ref{eq:ExplicitCFs}) add constructively for the intensity auto-correlation function, indicative of photon coalescence. Of course, the form of the intensity correlation functions for non-zero $\tau$ depends on the form of the mode functions, and we evaluate this now for Gaussian and Lorentzian photons. 

\subsection{Single Gaussian Photons}
One form of mode function commonly produced by single-photon sources are \emph{Gaussian} modes. For example, this is the case for trapped atom sources in which linewidth broadening mechanisms are dominant \cite{legero2}. In the subsequent section we consider the case of Lorentzian mode functions, which arise naturally for circuit QED systems. Given the recently demonstrated ability to shape the waveforms of single photons \cite{martinis:pulse}, we could consider more general mode functions as well. Two Gaussian modes, with a carrier frequency difference $\Delta$ (centred on $\omega_0$), a relative temporal offset $\delta \tau$ (centred on $t=0$), and a common pulse width specified by $\sigma$, are described by the spatio-temporal mode functions:
\begin{equation}
\zeta_{a(b)} (t) = \sqrt[4]{\frac{1}{\pi \sigma^2}} \exp \left[ - \frac{\left( t \pm \delta \tau /2 \right)^2}{2\sigma^2} - i \left( \omega_0 \pm \Delta /2 \right) t \right] . \label{eq:GaussianPhoton}
\end{equation}
%Of course, they are also Gaussian in the frequency domain.%, with the corresponding spectral densities, 
%\begin{eqnarray}
%\Phi_{a(b)} ( \omega ) & = & \sqrt[4]{ \frac{\sigma^2}{\pi} } \exp \left\{ \mp i\left[ \omega - \left( \omega_0 \pm \Delta/2 \right) \right] \delta \tau /2 \right\} \nonumber \\
%& & \times \exp \{ - \left[ \omega - \left( \omega_0 \pm \Delta/2 \right) \right]^2 \sigma^2/2 \} .
%\end{eqnarray}
Substituting Eq.~(\ref{eq:GaussianPhoton}) into Eq.~(\ref{eq:ExplicitCFs}) gives the explicit form of the output intensity correlation functions.
%\begin{eqnarray}
%\left\langle : \hat{n}_a (t+\tau ) \hat{n}_{a(b)} (t) : \right\rangle = \frac{ \cosh \left( \delta \tau \ \tau /\sigma^2 \right) \pm \cos \left( \tau \Delta \right) }{2\pi \sigma^2} \nonumber \\
%\times \exp \left\{ - \left[ 4t (t+\tau ) + \delta \tau^2 + 2\tau^2 \right]/ 2\sigma^2 \right\} ,
%\end{eqnarray}
Subsequently, integrating over the first detection time $t$ leads to the second-order correlation functions, 
\begin{eqnarray} 
G^{(2)}_{aa(ab)}(\tau ) & = & \frac{ \cosh \left( \tau \ \delta \tau /\sigma^2 \right) \pm \cos \left( \tau \Delta \right) }{ 2 \sqrt{2\pi \sigma^2} } \nonumber \\
& & \times \exp \left[ - \left( \delta \tau^2 + \tau^2 \right) / 2\sigma^2 \right] . 
\end{eqnarray}

The total \emph{correlation probabilities}, $P_{aa}$ and $P_{ab}$, are then obtained by integrating over the detection time $\tau$. The correlation probabilities $P_{cd}$ simply give the probability of detecting one photon in mode $c$ and one photon in mode $d$, for one pair of input photons, irrespective of the detection times. They are:
\begin{equation} 
P_{aa(ab)} =  \frac{1}{4} \left[ 1 \pm \exp \left( - \frac{\Delta^2 \sigma^2}{2} - \frac{\delta \tau^2}{2\sigma^2} \right) \right] . 
\end{equation}
Note that $P_{ba} = P_{ab}$ and $P_{bb} = P_{aa}$, such that $\sum_{c,d \in \{a,b\}} P_{cd} = 1$, as required for a beam splitter that is assumed to be lossless. The total probability for photon coalescence at one output port is $P_{aa}+P_{bb}=2P_{aa}$, while the total probability for one photon in either output port, the \emph{coincidence probability}, is $P_c \equiv P_{ab} + P_{ba} = 2P_{ab}$. The correlation probabilities may be regarded as probability distributions over the controlled parameters $\delta \tau$ and $\Delta$. For indistinguishable photons ($\delta \tau = 0$ and $\Delta = 0$) the coincidence probability is $P_c = 0$, and for fully distinguishable photons (i.e. $\left| \delta \tau \right| \gg \sigma$ and/or $\left| \Delta \right| \gg 1/\sigma$) we have $P_c \rightarrow 1/2$. The coincidence probability, plotted as a function of either $\delta \tau$ or $\Delta$, exhibits the classic \emph{Hong-Ou-Mandel dip} \cite{HOM}.

\subsection{Single Lorentzian Photons}
The same calculation may be performed for \emph{Lorentzian} photons. In the experiment of Ref.~\onlinecite{lang:microwaveHOM} the spatio-temporal mode function of the released photon is determined by the decay rate of the coupled circuit QED system. Accordingly, these photons have the shape of a truncated exponential decay. Two such mode functions, with a relative temporal offset $\delta t$ and a carrier frequency difference $\Delta$, are given by:
\begin{eqnarray}
\zeta_{a(b)}(t) & = & \sqrt{2\gamma} e^{-\gamma \left( t \pm \delta t /2 \right)} u(t \pm \delta t/2) e^{-i (\omega_0 \pm \Delta/2)t} , \nonumber \\
& & \label{eq:Lorentzian}
\end{eqnarray}
where $\gamma$ is the field decay rate of the coupled system, $\omega_0$ is the frequency about which the two carrier frequencies are centred, and $u(\ldots )$ denotes the unit step function. Note that the field decay rate is the characteristic decay rate of the electric field, in contrast to the commonly quoted (cavity) decay rate of the energy stored in the cavity. According to Eq.~(\ref{eq:Lorentzian}), the first photon (in mode $a'$) is ``released'' at $t=-\delta t/2$, while the second photon (in mode $b'$) is released at $t=+\delta t/2$. In the frequency domain, the mode functions correspond to the spectral densities,
\begin{eqnarray}
\Phi_{a(b)} ( \omega ) & = & \sqrt{ \frac{\gamma}{\pi} } \frac{1}{ \gamma - i \left[ \omega - \left( \omega_0 \pm \Delta/2 \right) \right] } \nonumber \\
& & \ \times \exp \left\{ - i \left[ \omega - \left( \omega_0 \pm \Delta/2 \right) \right] \delta t/2 \right\} .            
\end{eqnarray}
This representation makes it clear why the exponentially-decaying single-photon pulses described by Eq.~(\ref{eq:Lorentzian}) are referred to as Lorentzian photons. Of course, the mode function thus defined is unphysically sharp. The filtering effects of a finite measurement bandwidth shall be addressed below. 

Substituting Eq.~(\ref{eq:Lorentzian}) into Eq.~(\ref{eq:ExplicitCFs}), assuming that $\delta t > 0$ and $\tau > 0$, we find the output intensity correlation functions to be:
\begin{eqnarray}
& & \left\langle : \hat{n}_{a}(t+\tau ) \hat{n}_{a(b)}(t) : \right\rangle \nonumber \\
& = & \gamma^2 e^{-2\gamma (2t+\tau )} u(t+\tau-\delta t/2) u(t+\delta t/2) \nonumber \\
& & - \gamma^2 e^{-2\gamma (2t+\tau )} u(t-\delta t/2 ) \left\{ \begin{array}{l}  1 - 4 \cos^2 (\Delta \tau/2)  \\ 1 - 4 \sin^2 (\Delta \tau/2) \end{array} \right. , \nonumber \\ 
& & \label{eq:numberauto} 
\end{eqnarray}
where the upper (lower) result corresponds to the auto-correlation (cross-correlation) function. Eq.~(\ref{eq:numberauto}) consists of two contributions, corresponding to different combinations of the unit step functions switching on. The first product of step functions in each switches on when both the first detection ($t$) is after the release of the first photon ($-\delta t/2$) and the second detection time ($t+\tau$) is after the release of the second photon ($+\delta t/2$). The second unit step function is switched on when the first, and therefore the second detection time ($\tau > 0$) is after the release of the second photon. Integrating over the first detection time $t$ leads to the output second-order correlation functions,
\begin{eqnarray}
G^{(2)}_{aa(ab)}(\tau ) & = & \frac{\gamma}{2} e^{-2\gamma \delta t} \left[ \sinh 2\gamma \tau \ u (\delta t - \tau ) \right. \nonumber \\ 
& & \left. + 2e^{-2\gamma \tau} u (\delta t - \tau )  \left\{ \begin{array}{l} \cos^2 \left( \Delta \tau /2 \right) \\ \sin^2 \left( \Delta \tau /2 \right) \end{array} \right. \right] 
\nonumber \\
& & + \frac{\gamma}{2} e^{-2\gamma \tau} \left[ \sinh 2\gamma \delta t \ u(\tau - \delta t) \right. \nonumber \\
& & \left. + 2e^{-2\gamma \delta t} u(\tau - \delta t) \left\{ \begin{array}{l} \cos^2 \left( \Delta \tau/2 \right) \\ \sin^2 \left( \Delta \tau /2 \right) \end{array} \right. \right] . \nonumber \\
& & \label{eq:coherencecross} 
\end{eqnarray}
Again, Eq.~(\ref{eq:coherencecross}) is split into two terms, depending on whether the difference in measurement times ($\tau$) is greater or less than the temporal delay ($\delta t$) between the release of photons from either source. Integrating over the difference in detection times $\tau$ (positive and negative) we obtain the correlation probabilities,
\begin{equation}
P_{aa(ab)} = \frac{1}{4} \left( 1 \pm \frac{4 \gamma^2}{ 4\gamma^2 + \Delta^2 } e^{-2\gamma \delta t} \right) . 
\end{equation}
As before, we have $P_{bb}=P_{aa}$, $P_{ba}=P_{ab}$, $\sum P_{cd}=1$, and the coincidence probability is $P_{c} = P_{aa} + P_{bb} = 2P_{aa}$. Again, these probabilities may be regarded as probability distributions over $\delta t$ and $\Delta$.  If the photons are indistinguishable ($\delta t = 0$ and $\Delta = 0$), $P_c = 0$, while for fully distinguishable photons ($\delta t \gg 1/\gamma$ and/or $\left| \Delta \right| \gg \gamma$), $P_c \rightarrow 1/2$. 

\begin{figure*}[ht]
\begin{center}
\includegraphics[scale=0.5]{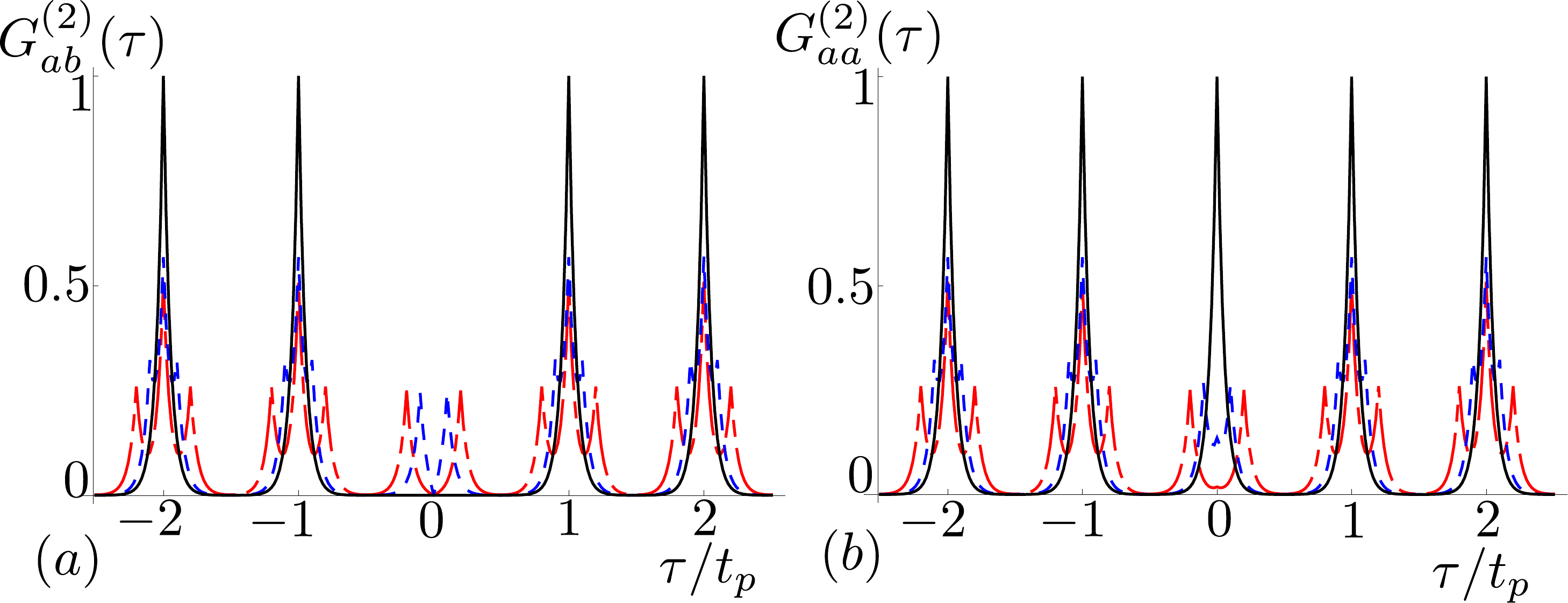}
\end{center}
\caption{ Output second-order (a) cross-correlation functions and (b) auto-correlation functions for trains of Lorentzian single photons input at each port of a beam splitter with pulse period $t_p$. The delay of one source with respect to the other in each case is $\delta t = 0 \ {\rm ns}$ (black line), $50 \ {\rm ns}$ (dashed blue line), $100 \ {\rm ns}$ (dashed red line) for $t_p = 0.5 \ {\rm \mu s}$. For indistinguishable photons (black line), the signal in the first pulse period of the cross-correlation function is suppressed, indicating interference.  } 
\label{fig:TrainG2}
\end{figure*}

\subsection{Trains of Single Photons}
The experiment in Ref.~\onlinecite{lang:microwaveHOM} is performed with \emph{trains} of pulsed single photons. In general, this can lead to features in the correlation functions at (and around) integer multiples of the pulse period. These are evaluated here. In order to determine the output intensity correlation functions in this case, we consider each photon as belonging to an independent spatio-temporal mode, and assume that the state of the many-photon system is separable with respect to this basis of spatio-temporal modes. Accordingly, the time-dependent quantum fields at the inputs may be decomposed as the sum of the fields corresponding to each spatio-temporal mode:
\begin{eqnarray}
\hat{a}' (t) = \sum^{N}_{k=-N} \zeta_{ak}(t) \hat{a}'_{0k}, \ \ \ \hat{b}' (t) = \sum^{N}_{k=-N} \zeta_{bk}(t) \hat{b}'_{0k} , \label{eq:decompositionb}
\end{eqnarray}
with the spatio-temporal mode operator commutation relations being $[ \hat{c}'_{0k}, \hat{c}'^\dagger_{0k'} ] = \delta_{kk'}$, where $\hat{c}'$ denotes an input mode lowering operator, $\hat{a}' $ or $\hat{b}' $. The appropriate spatio-temporal mode functions for the $k^{th}$ photon of each input (assumed Lorentzian here) are:
\begin{eqnarray}
\zeta_{ak(bk)}(t) & = & \sqrt{2\gamma } e^{-i \left( \omega_0 \pm \Delta/2 \right) t} \nonumber \\
& & \times e^{ -\gamma \left( t - kt_p \pm \delta t/2 \right) } u \left( t - kt_p \pm \delta t/2 \right) , 
\end{eqnarray}
with $t_p$ denoting the pulse period. Substituting the field decompositions of Eq.~(\ref{eq:decompositionb}) into the intensity correlation functions of Eq.~(\ref{eq:CFEvenMoments}) leads to:
\begin{widetext}
\begin{eqnarray}
\left\langle : \hat{n}_a (t+\tau ) \hat{n}_{a(b)} (t) : \right\rangle & = & \frac{1}{4} \sum_{(c,d) \in \Pi (a,b)} \left\{ \left\langle \left( \sum_k \zeta^*_{ck}(t) \hat{c}^\dagger_{0k} \right) \left( \sum_l \zeta^*_{cl}(t+\tau) \hat{c}^\dagger_{0l} \right) \left( \sum_m \zeta_{cm}(t+\tau) \hat{c}_{0m}  \right) \left(  \sum_n \zeta_{cn}(t) \hat{c}_{0n}   \right) \right\rangle \right. \nonumber \\
%& & + \frac{1}{4} \left\langle \left( \sum_k \zeta^*_{bk}(t) \hat{b}^\dagger_{0k} \right) \left( \sum_l \zeta^*_{bl}(t+\tau) \hat{b}^\dagger_{0l} \right) \left( \sum_m \zeta_{bm}(t+\tau) \hat{b}_{0m}  \right) \left(  \sum_n \zeta_{bn}(t) \hat{b}_{0n} \right) \right\rangle \nonumber \\
& & + \left\langle  \left( \sum_k \zeta^*_{ck}(t) \hat{c}^\dagger_{0k} \right) \left( \sum_l \zeta_{cl}(t) \hat{c}_{0l} \right) \right\rangle \left\langle  \left( \sum_m \zeta^*_{dm}(t+\tau) \hat{d}^\dagger_{0m} \right) \left( \sum_n \zeta_{dn}(t+\tau) \hat{d}_{0n} \right)   \right\rangle \nonumber \\
%& & + \frac{1}{4} \left\langle  \left( \sum_k \zeta^*_{ak}(t) \hat{a}^\dagger_{0k} \right) \left( \sum_l \zeta_{al}(t) \hat{a}_{0l} \right)   \right\rangle \left\langle  \left( \sum_m \zeta^*_{bm}(t+\tau) \hat{b}^\dagger_{0m} \right) \left( \sum_n \zeta_{bn}(t+\tau) \hat{b}_{0n} \right)   \right\rangle \nonumber \\
& & \left. \pm \left\langle  \left( \sum_k \zeta^*_{ck}(t+\tau) \hat{c}^\dagger_{0k} \right) \left( \sum_l \zeta_{cl}(t) \hat{c}_{0l} \right)   \right\rangle \left\langle  \left( \sum_m \zeta^*_{dm}(t) \hat{d}^\dagger_{0m} \right) \left( \sum_n \zeta_{dn}(t+\tau) \hat{d}_{0n} \right) \right\rangle \right\} . \label{eq:trainCF}
%& & \pm \frac{1}{4} \left\langle  \left( \sum_k \zeta^*_{ak}(t) \hat{a}^\dagger_{0k} \right) \left( \sum_l \zeta_{al}(t) \hat{a}_{0l} \right)   \right\rangle \left\langle  \left( \sum_m \zeta^*_{bm}(t+\tau) \hat{b}^\dagger_{0m} \right) \left( \sum_n \zeta_{bn}(t) \hat{b}_{0n} \right)   \right\rangle 
\end{eqnarray}
\end{widetext}
Using the assumption that the input state is separable with respect to the basis of spatio-temporal modes introduced, Eq.~(\ref{eq:trainCF}) reduces to 
\newline
\newline
\begin{widetext}
\begin{eqnarray}
\left\langle : \hat{n}_a(t+\tau ) \hat{n}_{a(b)}(t) : \right\rangle & = & \frac{1}{4} \sum_{(c,d) \in \Pi (a,b)} \left\{ \sum_m \left[ \zeta^*_{cm}(t) \zeta^*_{cm}(t+\tau ) \zeta_{cm}(t+\tau ) \zeta_{cm}(t) \left\langle \hat{c}^\dagger_{0m} \hat{c}^\dagger_{0m} \hat{c}_{0m} \hat{c}_{0m} \right\rangle \right] \right. \nonumber \\
& & \ \ \ \ \ \ \ + \sum_{m,n (m \neq n)} \left[ \zeta^*_{cm}(t) \zeta_{cm}(t+\tau ) \zeta^*_{cn}(t+\tau ) \zeta_{cn}(t) \left\langle \hat{c}^\dagger_{0m} \hat{c}_{0m} \right\rangle \left\langle \hat{c}^\dagger_{0n} \hat{c}_{0n} \right\rangle \right] \nonumber \\
& & \ \ \ \ \ \ \ \left. + \sum_{m,n (m \neq n)} \left[ \zeta^*_{cn}(t) \zeta_{cn}(t) \zeta^*_{cm}(t+\tau ) \zeta_{cm}(t+\tau ) \left\langle \hat{c}^\dagger_{0m} \hat{c}_{0m} \right\rangle \left\langle \hat{c}^\dagger_{0n} \hat{c}_{0n} \right\rangle \right] \right. \nonumber \\
%& & + \frac{1}{4} \left\{ \sum_m \left[ \zeta^*_{bm}(t) \zeta^*_{bm}(t+\tau ) \zeta_{bm}(t+\tau ) \zeta_{bm}(t) \left\langle \hat{b}^\dagger_{0m} \hat{b}^\dagger_{0m} \hat{b}_{0m} \hat{b}_{0m} \right\rangle \right] \right. \nonumber \\
%& & \ \ \ \ \ \ \ \ \ \ + \sum_{m,n (m \neq n)} \left[ \zeta^*_{bm}(t) \zeta_{bm}(t+\tau ) \zeta^*_{bn}(t+\tau ) \zeta_{bn}(t) \left\langle \hat{b}^\dagger_{0m} \hat{b}_{0m} \right\rangle \left\langle \hat{b}^\dagger_{0n} \hat{b}_{0n} \right\rangle \right] \nonumber \\
%& & \ \ \ \ \ \ \ \ \ \ \left. + \sum_{m,n (m \neq n)} \left[ \zeta^*_{bn}(t) \zeta_{bn}(t) \zeta^*_{bm}(t+\tau ) \zeta_{bm}(t+\tau ) \left\langle \hat{b}^\dagger_{0m} \hat{b}_{0m} \right\rangle \left\langle \hat{b}^\dagger_{0n} \hat{b}_{0n} \right\rangle \right] \right\} \nonumber \\
& & \ \ \ \ \ \ \ + \left( \sum_k \left[ \zeta^*_{ck}(t) \zeta_{ck}(t) \left\langle \hat{c}^\dagger_{0k} \hat{c}_{0k} \right\rangle \right] \right) \left(  \sum_l \left[ \zeta^*_{dl}(t+\tau ) \zeta_{dl}(t+\tau ) \left\langle \hat{d}^\dagger_{0l} \hat{d}_{0l} \right\rangle \right]  \right) \nonumber \\
%& & + \frac{1}{4} \left\{ \left( \sum_k \left[ \zeta^*_{ak}(t) \zeta_{ak}(t) \left\langle \hat{a}^\dagger_{0k} \hat{a}_{0k} \right\rangle \right] \right) \left(  \sum_l \left[ \zeta^*_{bl}(t+\tau ) \zeta_{bl}(t+\tau ) \left\langle \hat{b}^\dagger_{0l} \hat{b}_{0l} \right\rangle \right]  \right) \right\} \nonumber \\
& & \ \ \ \ \ \ \ \pm \left. \left( \sum_k \left[ \zeta^*_{ck}(t+\tau) \zeta_{ck}(t) \left\langle \hat{c}^\dagger_{0k} \hat{c}_{0k} \right\rangle \right] \right) \left(  \sum_l \left[ \zeta^*_{dl}(t ) \zeta_{dl}(t+\tau ) \left\langle \hat{d}^\dagger_{0l} \hat{d}_{0l} \right\rangle \right]  \right) \right\} . \label{eq:NumberCFExplicit} 
%& & \pm \frac{1}{4} \left\{ \left( \sum_k \left[ \zeta^*_{ak}(t) \zeta_{ak}(t+\tau) \left\langle \hat{a}^\dagger_{0k} \hat{a}_{0k} \right\rangle \right] \right) \left(  \sum_l \left[ \zeta^*_{bl}(t+\tau ) \zeta_{bl}(t ) \left\langle \hat{b}^\dagger_{0l} \hat{b}_{0l} \right\rangle \right]  \right) \right\} . \label{eq:NumberCFExplicit}
\end{eqnarray}
\end{widetext}
The output second-order correlation functions, $G^{(2)}_{aa(ab)}(\tau )$, are then obtained by integrating over the first detection time $t$, as in Eq.~(\ref{eq:numberCF}). 

These correlation functions are plotted in Fig.~\ref{fig:TrainG2} here and Figs.~2(c)-(f) of Ref.~\onlinecite{lang:microwaveHOM}. Note that we actually calculate \emph{circular} correlation functions, meaning that we evaluate these sums for a finite pulse train, and then wrap the time index around the pulse train. This means that the peak heights at later pulse periods will be at the same height as those at earlier pulse periods, rather than showing an artificial decay in peak height that would otherwise be produced due to the finite nature of the pulse train. Furthermore, the correlation functions are normalised such that the peaks at $\pm kt_p$ in $G^{(2)}_{ab}(\tau )$ are unity for indistinguishable photons input. 

As expected, the signature of interference in Fig.~\ref{fig:TrainG2} is the suppression of the output second-order cross-correlation function for small delays $\tau$ for indistinguishable photons $(\delta t = 0)$. As the distinguishability of the photons is increased through increasing $\delta t$, one clearly sees corresponding peaks in the first repeat period. The peaks in the correlation functions at $\pm kt_p$ ($k$ an integer) exhibit side peaks at $\pm \delta t$. This is easily understood: a given photon of each pair may be correlated will be correlated with the first or second photon of the subsequent pair, as well as with the first or second photon of the preceding pair. 

\subsection{Filtered Response}
\label{sec:filtering}

Integrating the intensity correlation function of Eq.~(\ref{eq:NumberCFExplicit}) results in second-order correlation functions with the very sharp features seen in Fig.~\ref{fig:TrainG2}, due to the abrupt switching of the single-photon pulses described by the unit step and exponential function in Eq.~(\ref{eq:Lorentzian}). However, such features are not observed in experimental data due to filtering in the detection scheme. To facilitate comparison with experimental data, we incorporate this filtering into our calculations. The filtering is performed on the time-dependent \emph{signals} obtained by the data acquisition scheme in Ref.~\onlinecite{lang:microwaveHOM}, denoted by $S_{a(b)}(t)$. The second-order correlation functions of the \emph{fields} are obtained from the \emph{filtered} version of these signals, $S^f_{a(b)}(t)$, given by the convolution integral, 
\begin{equation}
S^{f}_{a(b)} (t ) = \int^{+\infty}_{-\infty} S_{a(b)}(t') f (t - t' ) dt' ,
\end{equation}
where $f$ is a filter function in the time domain. Now the correlation functions of the output \emph{fields} of interest may be obtained from the correlation functions of these measured \emph{signals}, and so for the sake of calculations, we may implement the filtering directly on the spatio-temporal mode functions; that is,
\begin{equation}
\zeta^{f}_{a(b)k} (t ) = \int^{+\infty}_{-\infty} \zeta_{a(b)}(t') f (t - t' ) dt' . \label{eq:filterspatiotemporal}
\end{equation}
The filtering in Ref.~\onlinecite{lang:microwaveHOM} was implemented in two stages. In the first stage, a square window filter was employed to eradicate specific frequency components associated with the analog-to-digital conversion. A subsequent filtering stage, implemented as a finite-impulse-response Chebyshev filter \cite{DSP}, reduced other noise sources. For the purpose of calculations, Eq.~(\ref{eq:filterspatiotemporal}) was implemented only once, but with a filter function representative of that which was done experimentally. The intensity correlation functions are then given by Eq.~(\ref{eq:NumberCFExplicit}), evaluated using the filtered spatio-temporal mode functions of the form of Eq.~(\ref{eq:filterspatiotemporal}). The second-order correlation functions are then obtained by integrating over the first detection time $t$, as in Eq.~(\ref{eq:numberCF}), and are shown as the theory curves in Figs.~2(c)-(f) of Ref.~\onlinecite{lang:microwaveHOM}.

\section{Continuously-Driven Sources}
\label{sec:continuouslydrivensources}

\subsection{Introduction}
Rather than assuming the circuit QED systems are configured as pulsed single-photon sources, we can assume that they are continuously-driven and operated in the regime of resonant photon blockade \cite{resonantphotonblockade}. The output microwave field from each source will exhibit sub-Poissonian and anti-bunched photon statistics \cite{carmichael,kimbleSubPoisson}, leading to the possibility of observing Hong-Ou-Mandel interference at the beam splitter \cite{blatt}. Treating only the lowest transition of the coupled circuit QED system \cite{blais}, we consider the source simply as a driven quantum two-level system (TLS). The physics of each source is then essentially the physics of resonance fluorescence \cite{mollow,scully}, with additional dephasing included. Note that this TLS is in fact formed from the \emph{coupled} system, and is therefore \emph{not} the ``qubit'' typically employed for superconducting quantum information processing. 

We denote the TLS transition frequency by $\omega_c$ (where $c$ denotes the corresponding input mode, either $a'$ or $b'$), and assume that it is subject to driving by a \emph{resonant} monochromatic, large-amplitude microwave field. Each system is described, in a frame rotating with respect to $H_0 = (\hbar \omega_c /2) \sigma_z$, by the Hamiltonian $ H_I = \hbar \left( \Omega \sigma_+ + \Omega^* \sigma_- \right)/2 $ where $\Omega \equiv 2g\beta $ is the \emph{Rabi frequency}, with $\beta$ denoting the coherent amplitude of the driving field and $g$ being the Jaynes-Cummings coupling between the TLS and the driving mode \cite{blais}. The coupling of the TLS to its environment may be fully described, in the \emph{Markovian} \emph{white noise} approximation, by damping at a rate $\gamma_1$ and dephasing at a rate $\gamma_p$. Accordingly, the evolution of the TLS density operator $\rho$ may be given in the form of a Lindblad master equation \cite{quantumoptics} as:
\begin{equation} 
\dot{\rho} = - \frac{i}{2} \left[ \Omega \sigma_+ + \Omega^* \sigma_-, \rho \right] + \gamma_1 \mathcal{D}\left[ \sigma_- \right] \rho - \frac{\gamma_p}{4} \left[ \sigma_z, \left[ \sigma_z , \rho \right] \right] , \label{eq:TLSME}
\end{equation}
where $\mathcal{D}[ \hat{A} ] \equiv \hat{A}\rho \hat{A}^\dagger - \frac{1}{2} \hat{A}^\dagger \hat{A} \rho - \frac{1}{2} \rho \hat{A}^\dagger \hat{A}$ is the so-called \emph{dissipative superoperator}. Note that in this model each source is fully characterized by the parameters $\gamma_1, \gamma_p, \Omega,$ and $\omega_c$.

\subsection{Output Correlation Functions}
As in the case of pulsed sources, the quantities of interest are the second-order correlation functions of the microwave fields at the outputs of the beam splitter. The master equation of Eq.~(\ref{eq:TLSME}) facilitates the evaluation of correlation functions of TLS operators, and from the theory of atomic spontaneous emission \cite{scully}, we know that the output field will be proportional to the TLS lowering operator,
\begin{equation}
\hat{c}(t) \sim \hat{\sigma}^c_- (t), \label{eq:fieldatomic}
\end{equation}
where $c$ denotes the input mode $a'$ or $b'$. The output intensity correlation functions then follow from Eq.~(\ref{eq:evaluatedNumberCF}), to within a constant factor, by making the substitution in Eq.~(\ref{eq:fieldatomic}). Rather than explicitly evaluating the coefficient in Eq.~(\ref{eq:fieldatomic}), we shall ultimately calculate \emph{normalised} (field) correlation functions for which the coefficients will cancel. 

First we calculate the output correlation functions in terms of the input correlation functions. Suppose that the TLS transition frequencies are $ \omega_0 \pm \Delta/2$ (for the sources corresponding to the input modes $a'$ and $b'$, respectively), and that we are working in an interaction picture with respect to $H_0 = \sum_{c \in \{a',b' \} }(\hbar \omega_0/2) \sigma^c_z$. Then, from Eqs.~(\ref{eq:CFEvenMoments}) and (\ref{eq:fieldatomic}), the output intensity correlation functions (to within a constant factor) are:
\begin{eqnarray}
& &  \langle : \hat{n}_a (t+\tau ) \hat{n}_{a(b)} (t) : \rangle \nonumber \\
& & \sim \frac{1}{4} \sum_{(c,d) \in \Pi (a',b')} \left[ \langle \sigma^{c}_+ (t) \sigma^{c}_+ (t+\tau ) \sigma^{c}_- (t+\tau ) \sigma^{c}_- (t) \rangle \right. \nonumber \\
%& &  \ \ \ \left. + \langle \sigma^{b'}_+ (t) \sigma^{b'}_+ (t+\tau ) \sigma^{b'}_- (t+\tau ) \sigma^{b'}_- (t)  \rangle \right. \nonumber \\
& &  \ \ \ \ \ \ \left. + \langle \sigma^{c}_+(t) \sigma^{c}_-(t) \rangle \langle \sigma^{d}_+(t+\tau ) \sigma^{d}_-(t+\tau ) \rangle \right] \nonumber \\
%& &  \ \ \ \left. + \langle \sigma^{b'}_+(t) \sigma^{b'}_-(t) \rangle \langle \sigma^{a'}_+(t+\tau ) \sigma^{a'}_-(t+\tau ) \rangle \right]  \nonumber \\ 
& & \ \ \ + \frac{1}{4} \left[ \pm e^{i\tau\Delta } \langle \sigma^{b'}_+ (t+\tau ) \sigma^{b'}_- (t) \rangle \langle \sigma^{a'}_+(t) \sigma^{a'}_-(t+\tau ) \rangle \right. \nonumber \\
& & \ \ \ \ \ \ \left. \pm e^{-i\tau\Delta } \langle \sigma^{a'}_+ (t+\tau ) \sigma^{a'}_-(t) \rangle \langle \sigma^{b'}_+(t) \sigma^{b'}_-(t+\tau ) \rangle \right] . \nonumber \\
& & \label{eq:intensityatomic} 
\end{eqnarray}

%\begin{eqnarray}
%& &  \langle : \hat{n}_a (t+\tau ) \hat{n}_{a(b)} (t) : \rangle \nonumber \\
%& & \sim \frac{1}{4} \sum_{(c,d) \in \Pi (a',b')} \left[ \langle \sigma^{a'}_+ (t) \sigma^{a'}_+ (t+\tau ) \sigma^{a'}_- (t+\tau ) \sigma^{a'}_- (t) \rangle \right. \nonumber \\
%& &  \ \ \ \left. + \langle \sigma^{b'}_+ (t) \sigma^{b'}_+ (t+\tau ) \sigma^{b'}_- (t+\tau ) \sigma^{b'}_- (t)  \rangle \right. \nonumber \\
%& &  \ \ \ \left. \pm e^{i\tau\Delta } \langle \sigma^{b'}_+ (t+\tau ) \sigma^{b'}_- (t) \rangle \langle \sigma^{a'}_+(t) \sigma^{a'}_-(t+\tau ) \rangle \right. \nonumber \\
%& &  \ \ \ \left. \pm e^{-i\tau\Delta } \langle \sigma^{a'}_+ (t+\tau ) \sigma^{a'}_-(t) \rangle \langle \sigma^{b'}_+(t) \sigma^{b'}_-(t+\tau ) \rangle \right. \nonumber \\
%& &  \ \ \ \left. + \langle \sigma^{a'}_+(t) \sigma^{a'}_-(t) \rangle \langle \sigma^{b'}_+(t+\tau ) \sigma^{b'}_-(t+\tau ) \rangle \right. \nonumber \\
%& &  \ \ \ \left. + \langle \sigma^{b'}_+(t) \sigma^{b'}_-(t) \rangle \langle \sigma^{a'}_+(t+\tau ) \sigma^{a'}_-(t+\tau ) \rangle \right]  . \label{eq:intensityatomic} 
%\end{eqnarray}

\begin{figure*}[ht]
\begin{center}
\includegraphics[scale=0.5]{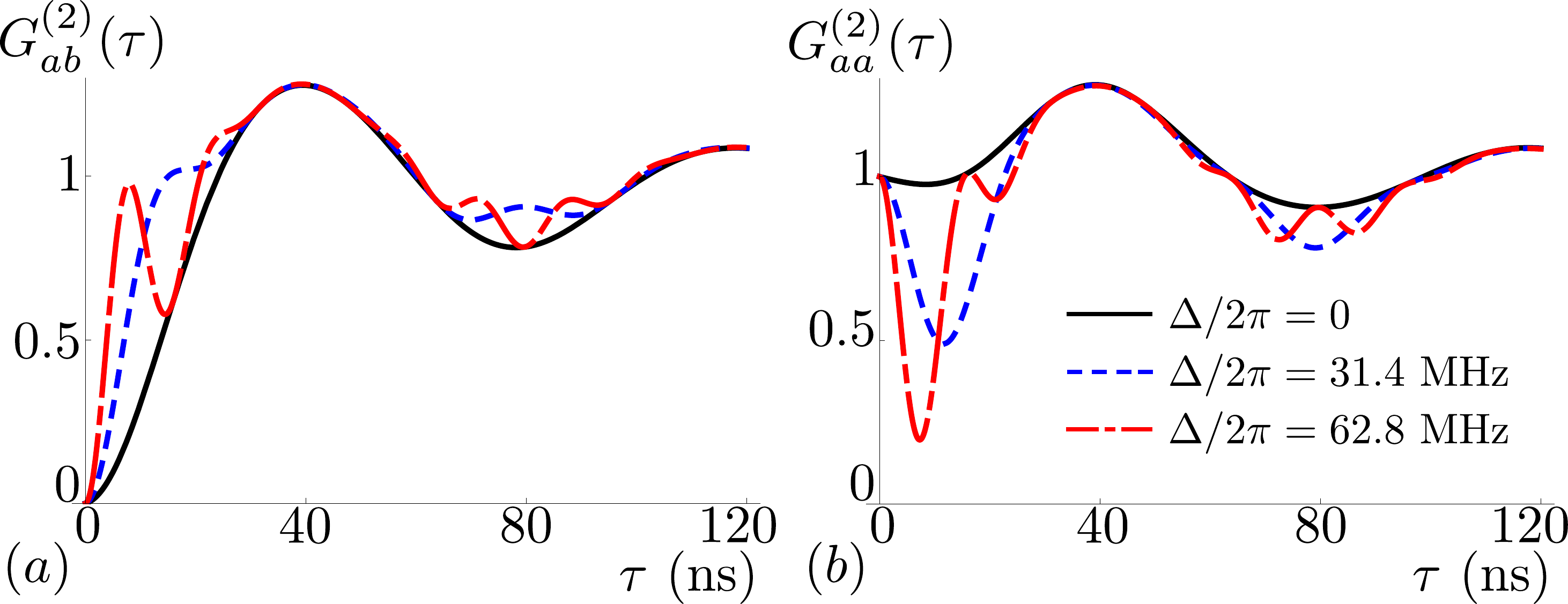}
\end{center}
\caption{ Output second-order (a) cross-correlation functions and (b) auto-correlation functions for each source continuously-driven in the regime of resonant photon blockade. Again, interference is observed in the suppression of the cross-correlation function for small $\tau$. Photon distinguishability may be introduced through a carrier frequency difference $\Delta$, which leads to an oscillation in the observed interference. } 
\label{fig:ResonanceFluorescenceG2}
\end{figure*}

Note that the validity of this expression depends on averaging out explicitly phase-dependent contributions by randomising the relative drive phases over repeated runs of the experiment, as discussed in Sec.~\ref{sec:phasedependentmoments}. Taking the limit $t \rightarrow \infty$ of Eq.~(\ref{eq:intensityatomic}) we have the output second-order correlation functions: 
\begin{eqnarray} 
G_{aa(ab)}(\tau ) & \sim & \frac{1}{4} \left\{ G^{(2)}_{a'} (\tau ) + G^{(2)}_{b'} (\tau ) + 2 G^{(1)}_{a'}(0) G^{(1)}_{b'}(0) \right. \nonumber \\
& & \ \ \ \left. \pm e^{i\tau\Delta } G^{(1)}_{a'} (\tau ) \left[ G^{(1)}_{b'} (\tau ) \right]^* \right. \nonumber \\
& & \ \ \ \left. \pm e^{-i\tau\Delta } \left[ G^{(1)}_{a'} (\tau ) \right]^* G^{(1)}_{b'} (\tau ) \right\} , 
\end{eqnarray}
where the correlation functions on the right-hand-side are now \emph{TLS} correlation functions, 
\begin{subequations}
\begin{eqnarray}
G^{(1)}_c (\tau ) & \equiv & \lim_{t \rightarrow \infty } \langle \sigma^c_+ (t) \sigma^c_- (t+\tau ) \rangle , \\ 
G^{(2)}_c (\tau ) & \equiv & \lim_{t \rightarrow \infty } \langle \sigma^c_+ (t) \sigma^c_+ (t+\tau ) \sigma^c_- (t+\tau ) \sigma^c_- (t) \rangle . \nonumber \\
& & 
\end{eqnarray}
\end{subequations}
%where $c$ denotes the corresponding input mode ($a'$ or $b'$). 
If one assumes that (carrier frequency aside) the sources have identical characteristics, meaning $G^{(2)}_{a'} = G^{(2)}_{b'} \equiv G^{(2)}_{i} $ and $G^{(1)}_{a'} = G^{(1)}_{b'} \equiv G^{(1)}_{i} $, then we have:
\begin{eqnarray}
G^{(2)}_{aa(ab)}(\tau ) & \sim & \frac{1}{2} \left\{ G^{(2)}_i (\tau ) \pm \cos \tau \Delta \left| G^{(1)}_i (\tau ) \right|^2 \right. \nonumber \\
& & \ \ \ \ \ \ \ \ \left. + \left[ G^{(1)}_i (0) \right]^2 \right\} . \label{eq:ctsCFs}
\end{eqnarray}
Again, the second and third terms in the output cross-correlation function of Eq.~(\ref{eq:ctsCFs}) cancel at $\tau = 0$. For non-zero $\tau$, the interference is oscillatory, an effect sometimes referred to as a \emph{quantum beat} \cite{legero2}.  

\subsection{Resonant Photon Blockaded Inputs}
Now we turn to the evaluation of the input TLS correlation functions. The system of equations for the expectations of TLS operators are readily calculated from Eq.~(\ref{eq:TLSME}). They are commonly referred to as the \emph{optical Bloch equations}, and may be expressed as the inhomogeneous system \cite{quantumoptics},
\begin{equation}
\frac{d}{dt} \left\langle \vec{\sigma} \right\rangle = A \left\langle \vec{\sigma } \right\rangle + \vec{b} , \label{eq:opticalBlochEquations}
\end{equation}
where we have introduced the vector of expectations of TLS (Pauli) operators, $ \left\langle \vec{\sigma} \right\rangle = \left( \left\langle \sigma_+ \right\rangle, \left\langle \sigma_- \right\rangle, \left\langle \sigma_z \right\rangle \right)^T $. The inhomogeneity is given by $\vec{b} = \left( 0,0,-\gamma_1 \right)^T$, and the system matrix is given by
\begin{equation}
A =  \left[ \begin{array}{ccc} - \gamma_2/2 + i \Delta & 0 & -i\Omega^*/2 \\ 0 & - \gamma_2/2 - i \Delta & i\Omega/2 \\ -i\Omega & i\Omega^* & - \gamma_1  \end{array} \right] ,
\end{equation}
where we have also introduced the \emph{total dephasing rate}, $\gamma_2 \equiv \gamma_1 + 2\gamma_p$. One can easily find the steady-state solution of Eq.~(\ref{eq:opticalBlochEquations}), $\left\langle \vec{\sigma} \right\rangle_{ss}$, and subsequently recast Eq.~(\ref{eq:opticalBlochEquations}) as the homogeneous system, $d \left\langle \vec{\sigma} \right\rangle'/dt = A \left\langle \vec{\sigma} \right\rangle'$ where $\left\langle \vec{\sigma} \right\rangle' \equiv \left\langle \vec{\sigma} \right\rangle - \left\langle \vec{\sigma} \right\rangle_{ss} $. Now the time-dependent solution of the homogeneous system may be obtained by diagonalization, and subsequently the solution to Eq.~(\ref{eq:opticalBlochEquations}) may be written in the form,
\begin{equation}
\left\langle \vec{\sigma} (t+\tau) \right\rangle = C(\tau ) \left\langle \vec{\sigma} (t) \right\rangle + \vec{d}(\tau ) , \label{eq:matrixsoln} 
\end{equation}
where we have introduced the notation
\begin{subequations}
\begin{eqnarray} 
C(\tau ) & \equiv & Xe^{ X^{-1} A X \tau } X^{-1} , \label{eq:Ctau} \\
\vec{d} (\tau ) & \equiv & (1 - Xe^{ X^{-1} A X \tau} ) \langle \vec{\sigma} \rangle_{ss} ,
\end{eqnarray}
\end{subequations}
with $X$ being the matrix of eigenvectors of $A$. 

The form of the solution in Eq.~(\ref{eq:matrixsoln}) is amenable to application of the \emph{quantum regression theorem} \cite{scully}. In particular, for an element of the solution in Eq.~(\ref{eq:matrixsoln}), we can write
\begin{equation}
\langle \sigma_i (t+\tau ) \rangle = \sum_j C_{ij} (\tau ) \langle \sigma_j (t) \rangle + d_i (\tau ) ,
\end{equation}
where $i \in \{ +,-,z \}$, the summation index $j$ runs over this set, and $C_{ij}(\tau )$ and $d_i(\tau )$ denote elements of the matrix $C(\tau )$ and vector $\vec{d} (\tau )$, respectively. The quantum regression theorem allows us to evaluate correlation functions as:
\begin{eqnarray}
\langle \sigma_k (t) \sigma_i (t+\tau ) \sigma_l (t) \rangle & = & \sum_j C_{ij} (\tau ) \langle \sigma_k(t) \sigma_j(t) \sigma_l(t) \rangle \nonumber \\
& & + d_i (\tau ) \langle \sigma_k(t) \sigma_l(t) \rangle . \label{eq:QRT}
\end{eqnarray}
The TLS correlation functions in Eq.~(\ref{eq:intensityatomic}) are obtained using Eq.~(\ref{eq:QRT}) and the algebraic properties of Pauli operators. 

According to Eq.~(\ref{eq:ctsCFs}), the output second-order correlation functions are fully determined by the input first-order and second-order correlation functions. We can write out $G^{(1)}_i (\tau )$ and $G^{(2)}_i (\tau )$, which are well-known results from the physics of resonance fluorescence \cite{scully}. For the first-order correlation functions we have: 
\begin{eqnarray}
G^{(1)}_i (\tau ) & = & \frac{\left| \Omega \right|^2}{\gamma_1 \gamma_2 + 2\left| \Omega \right|^2} \left[ \frac{\gamma^2_1}{\gamma_1 \gamma_2 + 2\left| \Omega \right|^2} + \frac{1}{2} e^{-\gamma_2 \tau/2} \right. \nonumber \\
& & \left. + \frac{1}{4} e^{-\left( 2\gamma_1 + \gamma_2 \right) \tau/4}  \left( \lambda_+ e^{ + \kappa \tau} + \lambda_- e^{ - \kappa \tau} \right)  \right] , \nonumber \\
& & \label{eq:G1ResonanceFluorescence}
\end{eqnarray}
where we have introduced the notation
\begin{subequations}
\begin{eqnarray} 
4\kappa & \equiv & \sqrt{ \left( 2\gamma_1 - \gamma_2 \right)^2 - 16 \left| \Omega \right|^2 } , \\
\lambda_\pm & = & \left. \left\{ 2 \left| \Omega \right|^2 - 2\gamma^2_1 + \gamma_1 \gamma_2 \pm \left[ 2 \left| \Omega \right|^2 \left( 6\gamma_1 - \gamma_2 \right) \right. \right. \right.\nonumber \\ 
& & \left. \left. \left. - \gamma_1 \left( 2\gamma_1 - \gamma_2 \right)^2 \right] /4\kappa \right\} \middle/ \left( \gamma_1 \gamma_2 + 2 \left| \Omega \right|^2 \right) \right. . \nonumber \\
& & 
\end{eqnarray}
\end{subequations}
Note that the usual resonance fluorescence spectrum follows from the result in Eq.~(\ref{eq:G1ResonanceFluorescence}), and in the strong driving regime ($\left| \Omega \right| \gg \gamma_1,\gamma_2$) we can obtain a simple analytical expression describing the so-called \emph{Mollow triplet} \cite{mollow,scully} feature in the spectrum. For the second-order correlation function we have:
\begin{eqnarray}
G^{(2)}_i (\tau ) & = & \left( \frac{\left| \Omega \right|^2}{\gamma_1 \gamma_2 + 2\left| \Omega \right|^2} \right)^2 \left[ 1 - e^{ -\left( 2\gamma_1 + \gamma_2 \right) \tau /4 } \right.  \nonumber \\
& & \left. \ \ \ \times \left( \cosh \kappa \tau + \frac{2\gamma_1 + \gamma_2}{4\kappa} \sinh \kappa \tau \right) \right] . \label{eq:G2ResonanceFluorescence}
\end{eqnarray}
It is more common to quote the normalised result, which is:
\begin{eqnarray}
g^{(2)}_i (\tau ) & \equiv & \frac{ G^{(2)}_i (\tau )  }{ \left[ G^{(1)}_i (0) \right]^2 } \nonumber \\
& = & 1 - e^{-\left( 2\gamma_1 + \gamma_2 \right) \tau /4} \nonumber \\
& & \ \ \ \ \times \left( \cosh \kappa \tau + \frac{2\gamma_1 + \gamma_2 }{4\kappa} \sinh \kappa \tau \right) . 
\end{eqnarray}
Of course, this is a normalised \emph{TLS} correlation function. Here we shall normalise the output field correlation functions by imposing the requirement that they asymptote to unity. 

Now the output second-order correlation functions are given by Eq.~(\ref{eq:ctsCFs}), using the results in Eqs.~(\ref{eq:G1ResonanceFluorescence}) and (\ref{eq:G2ResonanceFluorescence}). They are shown in Fig.~\ref{fig:ResonanceFluorescenceG2}. As expected, the interference effect is observed in the suppression of the cross-correlation function for small delays $\tau$, while introducing distinguishability through the frequency difference $\Delta$ leads to an oscillation in the observed interference for non-zero $\tau$. 

\subsection{Photon Polarisation}
In the continuously-driven case just considered, photon distinguishability can only be introduced through a difference in the carrier frequencies. An alternative approach, relevant in a three-dimensional circuit QED architecture \cite{paik}, is to introduce distinguishability through a \emph{polarisation} degree of freedom. To treat this problem, we decompose the time-dependent fields into their (horizontal and vertical) polarisation components:
\begin{eqnarray}
\hat{c} (t) & = & \hat{c}_h (t) + \hat{c}_{v} (t), \label{eq:polarisationDecomposition} 
\end{eqnarray}
where the subscripts $h$ and $v$ denote the horizontally and vertically polarised components, respectively, of the field. Generalizing the expression of Eq.~(\ref{eq:CFs}) to include polarisation indices, and then summing over these indices, we can write the output intensity correlation functions as:
\begin{eqnarray}
& & \left\langle : \hat{n}_a ( t + \tau ) \hat{n}_{c} (t) : \right\rangle \nonumber \\
& = & \sum_{ {k,l} \in \left\{ h,v \right\} } \left\langle \hat{c}^\dagger_k (t) \hat{a}^\dagger_l (t+\tau ) \hat{a}_l(t+\tau ) \hat{c}_k (t) \right\rangle , \label{eq:generalCF}
\end{eqnarray}
with $c$ representing either output mode ($a$ or $b$), as before.  

\begin{figure*}[bht]
\begin{center}
\includegraphics[scale=0.5]{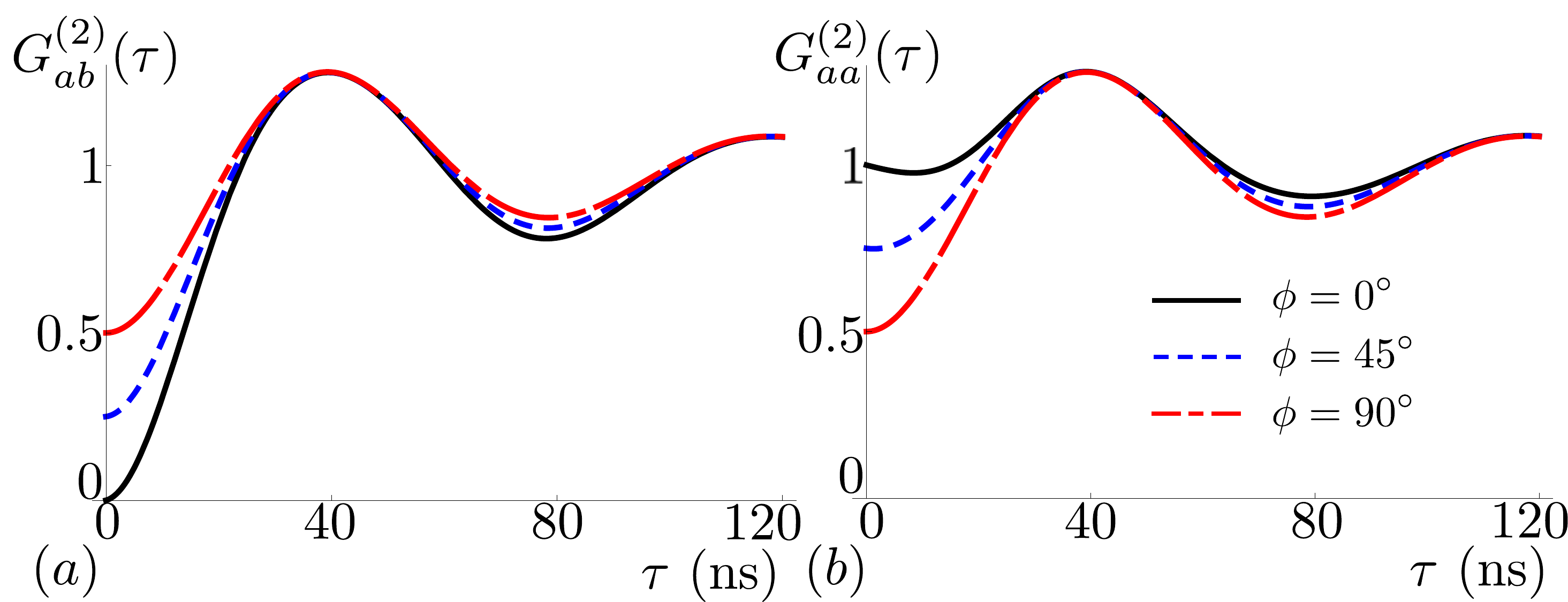}
\end{center}
\caption{ Output second-order (a) cross-correlation and (b) auto-correlation functions for each source continuously-driven in the regime of resonant photon blockade, for a range of polarisations of one input with respect to the other $(\phi = 0^\circ,45^\circ,90^\circ )$. Here the carrier frequencies of both inputs are assumed to be the same. For indistinguishable photons $(\phi = 0^\circ )$, the interference signature of a suppressed cross-correlation function for small $\tau$ is observed. As the indistinguishability (i.e. relative polarisation) is increased, the interference effect is reduced. }
\label{fig:PolarizationG2}
\end{figure*}

Now we take the components of the input fields to be related to the TLS lowering operators by:
\begin{subequations}
\begin{eqnarray}
\hat{a}'_h (t) & \sim & \sigma^{a'}_-(t) , \ \ \ \hat{a}'_v (t) = 0 , \\
\hat{b}'_h (t) & \sim & \sigma^{b'}_- (t) \cos \phi , \ \ \ \hat{b}'_v (t) \sim \sigma^{b'}_- (t) \sin \phi .
\end{eqnarray}
\end{subequations}
That is, the polarisation of one input with respect to the other is specified by the angle $\phi$. Accordingly, using Eq.~(\ref{eq:BS}) and the decomposition of Eq.~(\ref{eq:polarisationDecomposition}), we can find the polarisation components of the output fields. Substituting these into the generalised intensity correlation functions of Eq.~(\ref{eq:generalCF}) we find:  
\begin{eqnarray}
& & \left\langle : \hat{n}_a (t+\tau ) \hat{n}_{a(b)} (t) : \right\rangle \nonumber \\
& \sim & \frac{1}{4} \sum_{(c,d) \in \Pi ( a',b' ) } \left\{ \left\langle \sigma^c_+ (t) \sigma^c_+ (t+\tau ) \sigma^c_- (t+\tau) \sigma^c_- (t) \right\rangle \right. \nonumber \\
& & \left. + \left\langle \sigma^c_+ (t) \sigma^c_- (t) \right\rangle \left\langle \sigma^d_+ (t+\tau ) \sigma^d_- (t+\tau ) \right\rangle \right. \nonumber \\
& & \left. \pm \cos^2 \phi \left\langle \sigma^c_+ (t+\tau ) \sigma^c_- (t) \right\rangle \left\langle \sigma^d_+ (t) \sigma^d_- (t+\tau ) \right\rangle  \right\} . \label{eq:atomicpolarisation}
\end{eqnarray}
The photon distinguishability, and therefore the presence or absence of two-photon quantum interference, is determined by the relative polarisation angle $\phi$. For $\phi = 0$, the photons at either input are indistinguishable, and the second and third terms in Eq.~(\ref{eq:atomicpolarisation}) cancel for the cross-correlation function, corresponding to interference. For the auto-correlation function, the second and third terms will add constructively, indicative of photon coalescence. For $\phi = \pi/2$, the photons at either input are fully distinguishable, and the last term in Eq.~(\ref{eq:atomicpolarisation}) makes no contribution. Now taking the limit $t \rightarrow \infty$, as per Eq.~(\ref{eq:numberCF}), yields the second-order correlation functions, 
\begin{eqnarray}
G^{(2)}_{aa(ab)} ( \tau ) & \sim & \frac{1}{4} \sum_{(c,d) \in \Pi (a',b')} \left\{ G^{(2)}_{c}(\tau ) + G^{(1)}_{c}(0) G^{(1)}_{d}(0) \right. \nonumber \\
& & \ \ \ \left. \pm \cos^2 \phi \ G^{(1)}_{c} (\tau ) \left[ G^{(1)}_{d}(\tau )\right]^* \right\} .  
%\nonumber \\ & & \ \ \ \ \ \ \ \ \ \ \ \ \ \ \ \left. \left. + G^{(1)}_{b'}(\tau ) \left[ G^{(1)}_{a'}(\tau ) \right]^* \right) \right] .
\end{eqnarray}
Under the additional assumption that the two sources (polarisation aside) have identical properties, meaning $G^{(2)}_{a'} (\tau ) = G^{(2)}_{b'} (\tau ) \equiv G^{(2)}_{i} (\tau ) $ and $G^{(1)}_{a'} (\tau ) = G^{(1)}_{b'} (\tau ) \equiv G^{(1)}_{i} (\tau ) $, then we find:
\begin{eqnarray}
G_{aa(ab)}(\tau ) & \sim & \frac{1}{2} \left\{ G^{(2)}_i (\tau ) + \left[ G^{(1)}_i (0) \right]^2 \right. \nonumber \\
& & \ \ \ \ \ \ \ \left. \pm \cos^2 \phi \left| G^{(1)}_i (\tau ) \right|^2 \right\} . 
\end{eqnarray}
As for Eq.~(\ref{eq:atomicpolarisation}), the dependence of photon interference on the relative polarisation angle $\phi$ is clear. These output correlation functions are shown in Fig.~\ref{fig:PolarizationG2}. Clearly, access to a polarisation degree of freedom allows one to smoothly interpolate between the cases of distinguishable and indistinguishable photons in the continuously-driven case.  

\section{Conclusions}
Two-photon quantum interference at a beam splitter is observed in the suppression of the output second-order cross-correlation function for small detection delay times. In the case of pulsed single-photon sources, this quantity may be integrated over all detection delay times to reproduce the classic Hong-Ou-Mandel dip in the coincidence probability at the output. Accounting for a train of photons input leads to sidebands on the second-order correlation functions around integer multiples of the pulse period, while the inclusion of a finite detection bandwidth broadens the sharp features in the correlation functions that would otherwise be observed. The second-order correlation functions calculated are presented in Figs.~2(c)-(f) of Ref.~\onlinecite{lang:microwaveHOM}, showing very good agreement with the experimental data. In the case of continuously-driven sources, photon distinguishability may be introduced through a difference in carrier frequency or via a polarisation degree of freedom. In the former case, an oscillation in the quantum interference is expected. 

\section{ACKNOWLEDGEMENTS}
\noindent This work was supported by NSERC, CIFAR and the Alfred P. Sloan Foundation. C.L., C.E. and A.W. were supported by the European Research Council (ERC) through a Starting Grant and ETHZ. We thank Tom Stace, Andy Ferris and Clemens M\"{u}ller for valuable discussions.

\end{document}